\documentclass[prb,a4paper,superscriptaddress,floatfix,twocolumns]{revtex4}

\usepackage{graphics} 
 
\usepackage{latexsym} 
 
\usepackage{dcolumn}

\begin{document} 
 
 
\title{Effect of polymer-polymer interactions on the surface tension of 
colloid-polymer mixtures} 
 
\author{A.\ Moncho-Jord\'{a}} 
\email[Email:]{amjm3@cam.ac.uk} \affiliation{Department of Chemistry, Lensfield Road, Cambridge CB2 1EW, United Kingdom} 
 
\author{B.\ Rotenberg} 
 \affiliation{Department of Chemistry, Lensfield Road, Cambridge CB2 1EW, United Kingdom} 
 
\author{A.\ A.\ Louis} 
\email[Email:]{aal20@cus.cam.ac.uk} 
 \affiliation{Department of Chemistry, Lensfield Road, Cambridge CB2 1EW, United Kingdom}


\begin{abstract}

The density profile and surface tension for the interface of
phase-separated colloid-polymer mixtures have been studied in the
framework of the square gradient approximation for both ideal and
interacting polymers in good solvent. The calculations show that in
the presence of polymer-polymer excluded volume interactions the
interfaces have lower widths and surface tensions compared to the case
of ideal polymers. These results are a direct consequence of the
shorter range and smaller depth of the depletion potential between
colloidal particles induced by interacting polymers.

\end{abstract} 
 
 
\maketitle 
 
\section{Introduction} 
 
 Colloids are ``soft'' materials, readily deformable, with weak 
interfaces. This can be easily inferred from the ``giant atom'' 
picture of colloidal suspensions where, even though each colloid is 
made up of thousands of individual molecules, it is treated as a 
single particle interacting with an effective 
potential\cite{Frenkelbook,Liko01}. Since the effective interactions 
are roughly of the same shape as those of atomic fluids, an 
approximate corresponding states principle suggests that the reduced 
or dimensionless surface tensions should be similar.  Near the 
fluid-fluid transition, the attractive interactions for both classes of 
fluids are typically of order $k_B T$, but the colloidal particles 
have radii $R_{c}$ which can be $2$ or $3$ orders of magnitude larger than molecules. 
Thus the surface tension, which scales as $\gamma \sim k_B T/R_{c}^2$, 
is expected to be $4$ or more orders of magnitude lower than the 
values found for simpler atomic and molecular fluids. 
 Similar approximate corresponding states 
arguments also explain why colloidal crystals are so easily deformable: their 
elastic constants, which scale as $k_B T/R_{c}^3$, are at least 
$6$ orders of magnitude lower than those of simple atomic or 
molecular crystals.  Colloids are indeed a form of ``soft matter''. 
 
Surface tension plays an important role in the formation of
interfaces, as well as in phase transition kinetics, nucleation and
spinodal decomposition\cite{Barr03}.  Its indirect effects are
therefore easily observable, but its low values make direct
experimental measurements very difficult.  Nevertheless, some recent
experiments have made significant progress in measuring the
fluid-fluid interface of colloid-polymer mixtures and its surface
tension\cite{Hoog99,Hoog99_2}. In these systems, adding non-adsorbing
polymers induces attractive depletion pair potentials between the
colloids\cite{Asak54,Asak58}, which lead to the observed
phase-transition between a colloid-rich (``liquid'') and a
colloid-poor (``gas'') phase, separated by an interface.  Because the
experimental parameters can be easily tuned and controlled,
colloid-polymer mixtures form an important model system for the study
of phase transitions in soft matter\cite{Poon02}.
 
By applying theories similar to those used successfully for atomic 
and molecular fluids\cite{Rowl82}, Vrij\cite{Vrij97}, and  Brader 
and Evans\cite{Brad00}, calculated the properties of this fluid-fluid 
interface for the case of ideal polymers, finding qualitative 
agreement with experiments.  We have recently derived a depletion pair 
potential valid  for interacting polymers\cite{Loui02}, which captures 
the dominant effects of polymer-polymer interactions on the 
phase diagrams\cite{Rote03}.  This success suggests that the same 
potential can be used to calculate the properties of the fluid-fluid 
interface. 
 
The main purpose of this paper is to investigate the effects of
polymer-polymer interactions on the fluid-fluid interface of
colloid-polymer mixtures.  For that reason, we apply the same
combination of thermodynamic perturbation theory\cite{Hans86} and
square-gradient theory\cite{Rowl82} that was used by Brader and
Evans\cite{Brad00}, but with the new potential\cite{Loui02} instead of
the Asakura-Oosawa (AO)\cite{Asak54,Asak58} pair potential, valid only
for ideal polymers.  The differences between our new results, and
those of ref.~\cite{Brad00}, are then mainly due to the effect of
polymer-polymer interactions.

The use of colloid-colloid depletion pair potentials describes one
level of coarse-graining.  It is also possible to derive a more
fundamental two-component picture based on polymer-polymer,
polymer-colloid, and colloid-colloid pair potentials.
 A number of more recent investigations have used sophisticated
two-component density functional (DFT) theories for the AO
model\cite{Schm00} to uncover a host of interesting interfacial
phenomena, including oscillatory density profiles at the fluid-fluid 
interface and a series of layering transitions at the fluid-hard-wall 
interface\cite{Brad02}.  Computer simulations\cite{Dijk02} have
confirmed some of these results.  At present, all these theories are
only applicable to the AO model, and it is unfortunately not yet clear
how to extend them to interacting polymers (see however ref.~\cite{Schm03}) For that 
reason we restrict ourselves to the simplest square gradient approximation for
the interfacial profiles.
 
Our paper is organised as follows: after briefly reviewing the  
nature of the depletion potentials and the equilibrium phase-diagrams 
in section II, we describe the implementation of the square 
gradient approximation in section III, and present our results for the 
interfacial tension and width in section IV. Section V summarises our conclusions. 

\section{One-component effective depletion interactions} 

This section briefly describes the colloid-colloid 
effective depletion potentials for both ideal and interacting polymers.  
They are characterised by the polymer radius of gyration $R_g$, the colloid 
radius $R_c$, and the polymer number density $\rho_p$, or equivalently 
by the size-ratio $q = R_g/R_c$ and the reduced polymer density 
$\eta_p = \rho_p/\rho^*_p$, where $\rho^*_p = \frac43 \pi R_g^3$ is 
the so-called overlap density.  In the ideal case, the depletion 
interaction between two isolated colloidal spheres at distance \(r\) is 
accurately approximated by a potential of the Asakura-Oosawa form: 
\begin{eqnarray} 
\label{VAO} 
\beta V_{id}(r) = -\frac{4\pi}{3}\rho_p^r\sigma_{cp}^3 \left[ 
1-\frac{3}{4} \left( \frac{r}{\sigma_{cp}} \right) + 
\frac{1}{16}\left( \frac{r}{\sigma_{cp}} \right)^3 \right] 
\end{eqnarray} 
for \(2R_c < r < 2(R_c+R_{AO}^{eff})\); $V_{id}(r)=0$ for 
\(r>2(R_c+R_{AO}^{eff})\). Here, \(\sigma_{cp}=(R_c+R_{AO}^{eff})\) 
and \(\rho_p^r\) is the polymer density in a reservoir in osmotic equilibrium  
with the full colloid-polymer mixture\cite{Lekk92}.  
The range of this potential depends only on the 
polymer length and the depth is proportional to the polymer density. 
The effective Asakura-Oosawa parameter $R_{AO}^{eff}$ is set by the 
requirement that the insertion free energy of one colloid is equal to 
that of ideal polymers\cite{Meij94}; it is given by\cite{Loui02_116} 
\begin{equation} 
R_{AO}^{eff}=R_c \left[ \left( 1+\frac{6q}{\sqrt{\pi}}+3q^2 
\right)^{1/3} -1 \right]. 
\end{equation} 
For interacting polymers, we will use a recently proposed 
pair potential\cite{Loui02}, which accurately reproduces the depletion  
potentials obtained from direct computer simulations: 
\begin{equation} 
\label{Vs} 
V_s(r)=-\pi R_c\gamma_w(\rho_p^r)D_s(\rho_p^r)\left( 
1-\frac{r-2R_c}{D_s(\rho_p^r)} \right)^2 
\end{equation} 
for \(2R_c < r < 2R_c+D_s\) and $V_s(r)=0$ for 
\(r>2R_c+D_s\). Here \(\gamma_w(\rho_p^r)\) is the surface tension of the 
polymer solution near a single wall~\cite{Loui02_116} and 
\(D_s(\rho_p^r)\) is the range of the potential, given by 
\begin{equation} 
D_s(\rho_p^r)=\sqrt{\pi}\frac{\gamma_w(\rho_p^r)}{\Pi(\rho_p^r)} 
\frac{R_{AO}^{eff}}{R_g}, 
\end{equation} 
where \(\Pi\) is the osmotic pressure of the solution of interacting 
polymers, which is well understood\cite{Scha99}. 
 
The range of $V_{id}$ is independent of  density, whereas the range 
of $V_{s}$ shrinks with increasing density.  Furthermore, for a given 
$\rho_p^r$ and $R_g$, the well-depth of $V_{id}$ is greater than of 
$V_{s}$, which  implies that ideal polymers induce stronger depletion 
potentials than interacting polymers (see e.g.\ Fig.~2 of 
ref.~\cite{Rote03} for some explicit examples).  

 The differences in pair potentials help explain why, for a given $q$, 
phase-separation occurs at a larger value of $\eta_p$ for interacting 
polymers than for ideal polymers\cite{Rama02,Bolh02}, a difference 
that grows with increasing $q$.  Of course the pair-potential 
approximation becomes increasingly unreliable at high $q$ values, as 
many-body interactions become more important.  However, we have 
recently shown\cite{Rote03} that calculations based on pair potentials 
alone remain remarkably accurate up to $q \sim 1$.  In this paper 
we use the phase-diagrams calculated in ref.~\cite{Rote03}, based on 
second order perturbation theory, as the basis for our calculations of 
the properties of the fluid-fluid interface.  We make the implicit 
assumption that the effective Hamiltonian used for phase-behaviour is 
also appropriate for describing the interface. This follows from the fact 
that we are working at a contact chemical potential, so that the same 
effective potential holds across the density inhomogeneity occurring at 
the ``free'' interface (see ref.~\cite{Brad01}). 

\section{Interfacial properties from the square gradient approximation} 

Once phase separation occurs, there are two phases with well 
defined colloidal bulk densities (\(\rho_c^G\) and \(\rho_c^L\) for 
the dilute and concentrated colloidal phases, respectively). Both 
phases are separated by a planar interface where the local density 
depends on the distance to the interface, \(\rho_c(z)\).  
A well known way to treat the free-energy cost of making an interface 
is given by the square gradient approximation, where the free-energy 
is expanded  to lowest non-trivial order in a gradient expansion around 
the homogeneous fluid.  The surface tension and the density 
profile are then obtained from the integral of the free-energy across 
the interface\cite{Cahn58,Evan79,Rowl82,Evan92} 
\begin{equation}\label{eq3.1} 
\gamma=\int_{-\infty}^{\infty} \left[ \Psi \left(\rho_c(z) 
\right)+\kappa\left( \frac{d\rho_c}{dz} \right)^2 \right]dz ,
\end{equation} 
where \(\Psi\left( \rho_c(z) \right)=f\left( \rho_c(z)\right)-\mu_c 
\rho_c(z)+P\).  Here \(\mu_c\) and \(P\) are the chemical potential 
and osmotic pressure of the colloids at coexistence and \(f \left( \rho_c(z) \right) \)  
is the Helmholtz free energy density of a hypothetical colloid fluid of 
density \(\rho_c(z)\). The coefficient of the square gradient term, 
$\kappa$, describes the free-energy penalty for creating an interface. 
Minimising this functional\cite{Evan79,Rowl82} leads to the following 
expressions for the density profile 
\begin{equation} 
\label{rhoz} 
\left( \frac{d\rho_c}{dz} \right)^2 = \frac{\Psi}{\kappa} 
\end{equation} 
and the surface tension
\begin{equation} 
\label{gamma} 
\gamma = 2\int_{\rho_c^G}^{\rho_c^L}\left[ \kappa \Psi 
\right]^{1/2}d\rho_c. 
\end{equation} 
 
Requiring the functional in Eq.~(\ref{eq3.1}) to satisfy linear response 
relates the coefficient $\kappa$ to properties of the direct 
correlation function $c(r)$ of the {\em homogeneous} fluid
\cite{Rowl82} 
\begin{equation}\label{kappa1} 
\kappa=\frac{\pi k_BT}{3}\int_{0}^{\infty}r^4c(r,\rho_c)dr.
\end{equation}  

Note that all these variables depend implicitly on the polymer
chemical potential (or equivalently the polymer reservoir density,
$\rho_p^r$) of the corresponding coexistence point. Due to the factor
\(r^4\) in the integrand of expression~(\ref{kappa1}), the value of
\(\kappa\) is mainly determined by the behaviour of \(c(r)\) at large
\(r\), where it is well known that \(c(r) \approx -\beta V(r)\). We
therefore follow Ref. \cite{Brad00}, and set \(c(r,\rho_c)\) to be
zero for \(r<2R_c\) and equal to \(-\beta V(r)\) for \(r>2R_c\), where
\(V\) is given by Eqs.~(\ref{VAO}) and (\ref{Vs}) for ideal and
interacting polymers, respectively. Hence Eq.~(\ref{kappa1}) reduces to
\begin{equation} \label{kappa} 
\kappa \approx -\frac{\pi}{3}\int_{2R_c}^{\infty}V(r)r^4dr.
\end{equation}

This approximation has the further advantage that it circumvents the
conceptual difficulty of defining $c(r;\rho_c)$ in the coexistence
region. Even though the approximation for $c(r)$ itself may not always
be so reliable, we found that the values of $\kappa$ still compare
well with more sophisticated calculations of $c(r)$, because this
simple model interpolates between the values at the two coexistence
points.  Similar conclusions were reached in a paper studying Lennard
Jones systems\cite{Lu85}, where the Percus-Yevick approximation for
the low and high density fluid phases was combined with a lever rule
to obtain $c(r)$ in the coexistence region.

\section{Results and discussion} 

\subsection{Phase behaviour and interfacial properties} 

In our earlier work\cite{Rote03}, the free energy densities for the
effective one-component system, \(f(\rho_c)\), were calculated for
various \(q\)-ratios by second order perturbation theory using the
Barker-Henderson formulation \cite{Bark67}. The phase diagrams were
determined by the common tangent construction.  The resulting
coexistence curves for ideal and interacting polymers are plotted in
Fig.~\ref{Fig1}, for size-ratio \(q=0.67\), as a function of the
colloid packing fraction \(\eta_c=4\pi\rho_c R_c^3/3\) and polymer
reservoir packing fraction \(\eta_p^r=4\pi\rho_p^r R_g^3/3\). The
fluid-fluid coexistence lines are at higher values of polymer packing
fraction for interacting polymers, which implies that polymer excluded
volume effects reduce the global attraction between colloidal
particles. But not only is the position of the binodal different, so
is its shape. In particular, the binodal is flatter and the separation
between the critical and triple points is smaller for interacting
polymers.  This effect is not merely an artifact of the pair-potential
approximation, since the qualitative difference in shape is also
observed when comparing two-component simulations of ideal and
interacting polymer models\cite{Bolh02}.  Although we only show
results for one size-ratio in Fig.~\ref{Fig1}, the differences become
more pronounced for increasing $q$\cite{Bolh02,Rote03}, and are
finally quite dramatic in the so called ``protein limit'' where $q >>
1$\cite{Bolh03}.

It should be noted that, at least within our perturbation theory
treatment, the gas-liquid binodal obtained for interacting polymers at
\(q=0.34\) is metastable with respect to the fluid-solid
coexistence. This is not, however, an obstacle to the calculation of
surface tensions. Furthermore, in many experimental systems the
fluid-solid nucleation rates are very low, allowing the observation of
metastable fluid-fluid phase-separation.

The free energy densities from perturbation theory were used in
Eqs.~(\ref{rhoz}) and (\ref{gamma}) to calculate first the density
profiles and then the surface tensions for the coexistence points
along the fluid-fluid binodal. Two typical density profiles
corresponding to ideal and interacting polymers are shown in
Fig.~\ref{Fig2} for \(q=1.05\) and
\(\Delta\eta_p^r=(\eta_p^r-\eta_p^{r,crit})/\eta_p^{r,crit}=0.2\). All
profiles obtained using the square gradient theory share approximately
the same shape, i.e. a smooth monotonic curve which goes from the
dense to the dilute colloidal phase. Comparison between both curves
shows that the difference between the colloidal packing fractions in
the two phases is larger for interacting polymers (a consequence of
the flatter binodals), whereas the interfacial thickness is smaller.

The interfacial profiles are essentially characterised by their
width. The 10-90 width of the interface (\(W\)), defined as the
distance along the interface over which the colloidal density varies
from \((\eta_c^G+0.1(\eta_c^L-\eta_c^G))\) to
\((\eta_c^G+0.9(\eta_c^L-\eta_c^G))\), is plotted in Fig.~\ref{Fig3}
w.r.t.\ the deviation of the polymer reservoir packing fraction from
the critical point. As expected, the interface is very diffuse near
the critical point, and becomes sharper upon approaching the triple
point. The interfacial widths increase with increasing $q$, reflecting
the longer ranged attractions. For a given value of $q$, the widths
are consistently lower for interacting polymers than for ideal
polymers.

The resulting dimensionless surface tensions (\(\gamma R_c^2/k_BT\))
are shown in Fig.~\ref{Fig4} versus the difference in colloidal
packing fractions between both phases for three different size ratios,
\(q=0.34\), 0.67 and 1.05. For both ideal and interacting polymers,
the surface tension vanishes at the critical point, and increases with
\(\Delta\eta_c=\eta_c^L-\eta_c^G\) as expected.  These reduced values
are of the same order as those found for simple
liquids\cite{Rowl82}. Their absolute values depend only on $R_c$;
typical colloid sizes result in values of $\gamma$ near the triple
point on the order of $\mu N/m$ which is much smaller than the values
of $10 - 100 m N/m$ found for simple fluids.  For a given $\Delta
\eta_c$, the surface tension increases with $q$.  Moreover, for a
given $q$ the ideal polymer surface tension is always significantly
larger than that of interacting polymers.

In view of the mean field nature of the present theory, the critical
exponents are obviously classical, i.e. \(\gamma \propto
\Delta\eta_c^3\) and \(W \propto \Delta\eta_c^{-1}\). However, the
prefactors are expected to be different for ideal and interacting
polymer depletants. These prefactors are determined in the following
sub-section.  For ideal polymers similar scaling laws with the
reservoir density $\rho_p^r$ can also be derived, by exploiting the
analogy with inverse temperature.  For example\cite{Brad00}, \(W \propto
(\Delta\eta_p^r)^{-\frac12}\). However, for interacting
polymers such a simple scaling with $\rho_p^r$ does not follow,
because the effective well-depths (i.e.\ the inverse "temperatures")
don't scale in a simple way with this variable.

\subsection{Connection to the form of the depletion potential} 

The results of the previous subsection show that both the the 10-90
width, and the interfacial tension, are lower in the case of
interacting polymers. In this section, we will attempt to rationalise
these differences on the basis of the effective pair potentials.

Close to the critical point, \(\Psi(\eta_c)\) can be approximated as
the product of two quadratic potential wells centred around the
coexistence points\cite{Barr03}
\begin{equation} \label{psi*}
\Psi^{*}(\eta_c) \approx \frac{C}{2}\left( \eta_c-\eta_c^G
\right)^2\left( \eta_c-\eta_c^L \right)^2,
\end{equation}
where \(\Psi^{*}=(4\pi R_c^3/3)\beta\Psi\) is dimensionless and
\(C=(1/12)d^4\Psi^{*}/d\eta_c^4\), calculated at the critical point.
If we assume that $\kappa$ is independent of $\rho_c$, as we did in
the previous section, then inserting Eq.~(\ref{psi*}) into
Eq.~(\ref{gamma}) yields the following expression for the reduced
surface tension: 
\begin{equation} \label{gammaC}
\gamma^{*}=\frac{R_c^2\gamma}{k_BT}\approx
0.0275\sqrt{C}\sqrt{\kappa^{*}(\eta_p^r)}\left( \eta_c^L-\eta_c^G
\right)^3
\end{equation}
where \(\kappa^*=\beta\kappa/R_c^5\). The
dependence of \(\kappa^{*}\) on the polymer reservoir packing fraction
is indicated explicitly to remind us that it is not a constant
parameter but rather increases as we move away from the critical point
since it depends on \(\eta_p^r\) through \(V(r)\).  The same arguments
lead to\cite{Barr03} 
\begin{equation} 
W\sim \sqrt{\frac{\kappa^{*}(\eta_p^r)}{C}}\left( \eta_c^L-\eta_c^G
\right)^{-1} .
\end{equation}

The values of $W$ and $\gamma^*$ are determined by the 
parameters \(\kappa^{*}\) and \(C\). Eq.~(\ref{kappa})
directly relates \(\kappa\) to the depletion potential, whereas \(C\)
is given by the shape of the free energy inside the van der Waals
loop. Understanding how the depletion potentials govern the
interfacial behaviour now reduces to explaining how these two
parameters depend on the potentials.

Fig.~\ref{Fig5} compares the depletion pair potential for ideal and
interacting polymers at their corresponding critical points when
$q=0.67$. Even though the depletion potential for interacting polymers
has a larger depth at contact, its range is significantly shortened by
the polymer excluded volume interactions\cite{Loui02}. The integrand in
Eq.~(\ref{kappa}), which determines \(\kappa\), multiplies the potential
by $r^4$, giving extra weight to the longer ranged parts of the
potentials. Thus the ratio of $\kappa^*_{int}$ (calculated with
Eq.~(\ref{Vs})) to $\kappa^*_{id}$ (calculated with the potential of
Eq.~(\ref{VAO})), is always less than $1$, and decreases with increasing
$q$, as shown in the inset of Fig.~\ref{Fig5}. It has been recently
shown that the reduced second virial coefficient, which is
proportional to the integral of $r^2 (\exp[-\beta V(r)]-1)$, is very
similar at the critical point
 for a wide class of
attractive potentials\cite{Vlie00}. This observation remains true for the two
depletion potentials, giving further support to our argument that it
is the large factor $r^4$ in the integrand for $\kappa$ which is
responsible for the differences between $\kappa^*_{id}$ and
$\kappa^*_{int}$.

 The value of the other parameter, \(C\), is not as easy to link
directly to the pair potential.  Nevertheless, it can be determined
numerically from our previous calculations of the surface tension.  It
turns out not to depend strongly on \(q\).  We find \(C=85.8 \pm 0.5\)
for ideal and \(C=76.0 \pm 0.3\) for interacting polymers, the small
difference perhaps reflecting the fact that the free energy loop is
slightly flatter for interacting polymers.

Compared to the large changes in the surface tension, the values of
$C$ are quite similar, so that differences in the $\gamma^*$s, which
scale as \(\gamma\sim\sqrt{\kappa^{*} C}\), arise mainly from
$\kappa$. Since \(\kappa\) increases with the range of the potential,
i.e. with \(q\), this explains why, for a given type of depletant, the
surface tension grows with $q$.  It further shows how the main
differences in $\gamma^*$ between ideal and interacting polymers are
linked to the reduction of the depletion potential range induced by
polymer-polymer interactions. Since this effect becomes more important
for increasing $q$, we expect the differences between the surface
tensions induced by ideal and interacting polymers to grow with $q$ as
well, and to become more pronounced in the protein limit\cite{Bolh03},
as was recently pointed out by Sear\cite{Sear02,Sear02a}.

The interfacial width scales as \(W \sim \sqrt{\kappa^{*}/C}\). 
Again, differences in $W$ are dominated by $\kappa$.  Thus, the 
net effect of adding polymer-polymer interactions is to decrease the 
interfacial width. 

These results show that the decrease of the range in the depletion
potential, caused by the polymer interactions, plays the dominant role
in determining the differences in the interfacial properties between
the two types of depletants. The change in the depth of the potential
is only a secondary effect.

In all the arguments above it should be kept in mind that our double
symmetric parabola approximation of Eq.~(\ref{gammaC}) is only valid
close to the critical point.  For coexistence points far from it, the
bulk correlation length (defined as
\(\xi_b=(2\kappa/(d\mu_c(\rho_b)/d\rho_b))^{1/2}\), where \(\rho_b\)
is the colloidal density in the bulk phase) does differ between the
two phases, implying that the decay of the profile tails is different
between the liquid and the gas colloidal phase\cite{Lu85}.  Strictly
speaking, more sophisticated theories are needed in order to describe
the interfacial properties near the triple point.  DFT theories of the
fluid-fluid interface in the two-component AO model suggest that for
ideal polymers, the square gradient approximation underestimates the
values of the interfacial tension\cite{Brad02}, and that it misses
more subtle effects like oscillatory density profiles.  Part of the
difference come from using more accurate DFT's, but some also arises
from the fact that the two-component DFT yields different
phase-diagrams than the effective one-component
description\cite{Schm00}. In particular, the two-component theories
yield a larger distance between the critical point and the triple
point\cite{Meij94,Dijk02}, which may have an important qualitative and
quantitative effect on the interfacial behaviour.  While it would
clearly be desirable to have a two-component theory of similar
accuracy for interacting polymers, this is not available at present.
However, it is possible to use more sophisticated DFT approaches to
study the one-component interacting polymer system\cite{Wink01}, a
direction which we are now pursuing.  In fact, we have already
performed preliminary calculations using an accurate DFT for the HS
part, and with the interaction of Eq.~(\ref{Vs}) treated as a
perturbation.  The results for surface tensions and interfacial widths
are slightly higher than those from the present treatment, but the
trends are very similar. This suggests that our use of the square
gradient theory, coupled with our rater simple approximation for
$c(r)$, as used in Eq.~(\ref{kappa}), is quite reliable.

 In the longer term, it would be
interesting to develop some approximate density functional for
interacting polymers using a two-component representation, perhaps
along the lines of ref.~\cite{Schm03}, in order to obtain accurate
predictions of the interfacial properties of free fluid-fluid
interfaces, adsorption and wetting behaviour at hard walls and even
surface tensions and density profiles of fluid-solid interfaces.

Although the actual values of the surface tensions found within the
square gradient theory may only be accurate to about a factor of two,
the differences between ideal and interacting polymers are fairly
large, and follow from a simple physical explanation which seems
robust.  We therefore don't expect more sophisticated theories to
reverse the trends discussed in this paper. On the other hand, whether
or not the more subtle interfacial phenomena found for the
two-component AO model\cite{Brad02,Dijk02} will be even qualitatively
similar for interacting polymers remains to be seen.  For example, in
the latter case the differences between the triple and critical points
are less pronounced, which may lead to less well defined oscillations
in the density profiles.  Clearly more work remains to be done.

The square gradient theory does not include the effects of capillary
fluctuations\cite{Rowl82}, but these are not expected to be large, as
shown in ref.~\cite{Brad00}.  We can therefore make comparisons with
experiments.  Full two-component AO model calculations within
DFT\cite{Brad02} lead to values of the surface tension that are close
to those of recent experiments~\cite{Hoog99}. However, our results
suggest that including polymer-polymer interactions will lower the
value of the surface tension, leading to less agreement.

\section{Conclusion} 

We have used thermodynamic perturbation theory and the square-gradient
approximation to calculate the properties of the fluid-fluid interface
for mixtures of colloids and interacting polymers within an effective
one-component representation.  We find significant differences
compared to the case of ideal polymers. The main effect of
polymer-polymer excluded volume interactions is to reduce the value of
the interfacial tension $\gamma$ and the interface width $W$.  This
effect becomes more pronounced as the size-ratio $q$ increases. It can
be rationalised by the differences in depletion potentials: at the
critical point, the range for the interacting polymer case is
significantly less than for the ideal case.  This has a pronounced
effect on the square gradient prefactor $\kappa$, and helps explain
the differences between the two types of depletants.

\section{Acknowledgements} 

A. Moncho-Jorda thanks the Ram\'{o}n Areces Foundation (Madrid), 
B. Rotenberg thanks the Ecole Normale Superieure (Paris), and A.A. Louis 
thanks the Royal Society (London), for financial support.  We are indebted 
J. Dzubiella, C.N. Likos and R. Roth for helpful discussions and R. Evans 
and J. P. Hansen for a critical reading of the manuscript.

\newpage

A. Moncho-Jord\'{a} et al., Fig. 1.

\begin{figure} \label{Fig1}
\center\resizebox{1.0\textwidth}{!}{\includegraphics{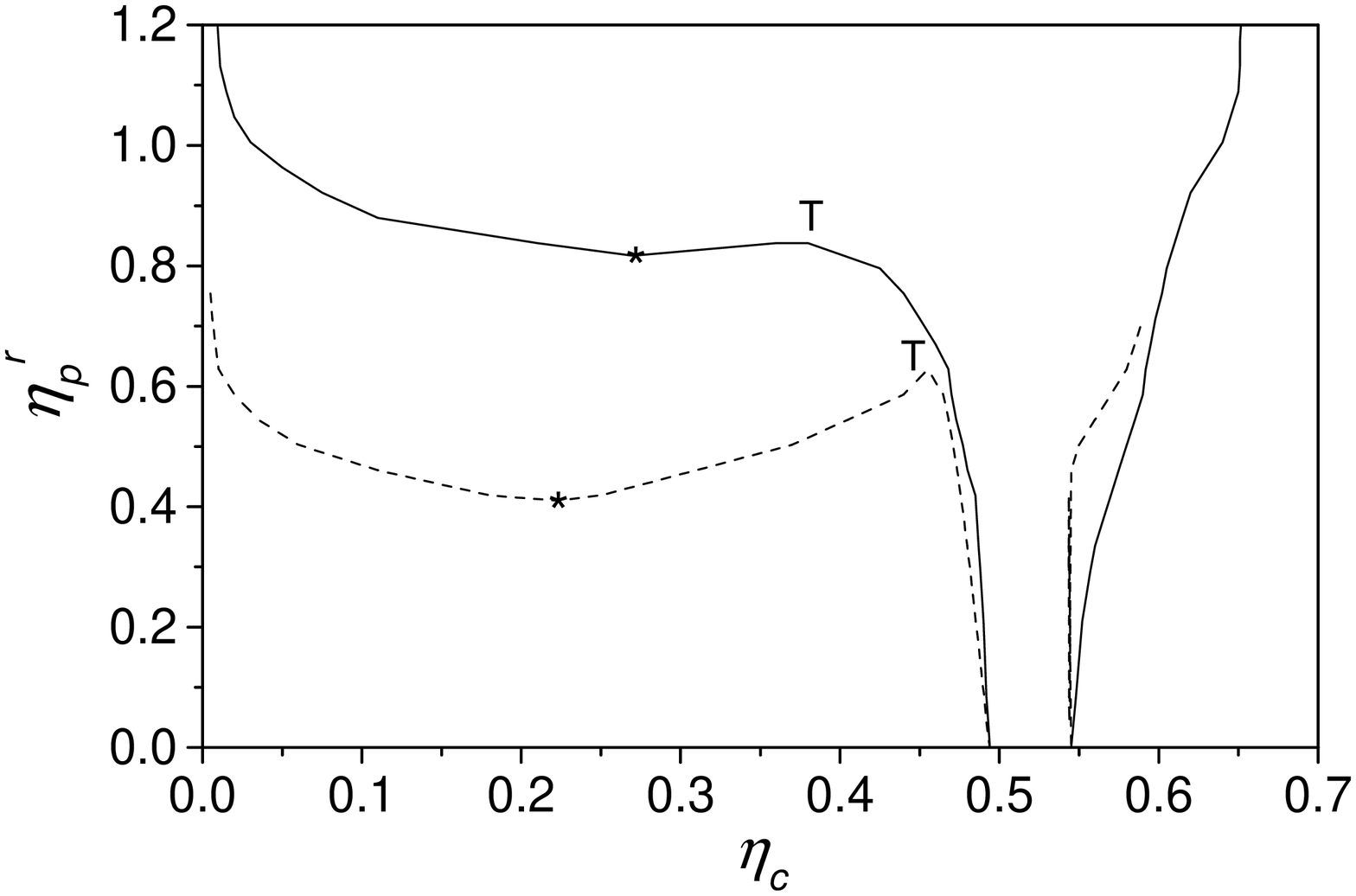}}
\caption{\label{Fig1} Phase diagrams of a colloid-polymer mixture for 
ideal (dashed lines) and interacting polymers (solid lines) compared for  
\(q=R_g/R_c=0.67\) (taken from \protect\cite{Rote03}). $\eta_c$ is the colloidal packing fraction and 
\(\eta_p^r\) the reservoir polymer packing fraction.  The binodal for 
interacting polymers is at a higher $\eta_p$ than that 
of ideal polymers, because the latter are stronger depletants than the 
former.}
\end{figure} 

\newpage

A. Moncho-Jord\'{a} et al., Fig. 2.

\begin{figure}
\center\resizebox{1.0\textwidth}{!}{\includegraphics{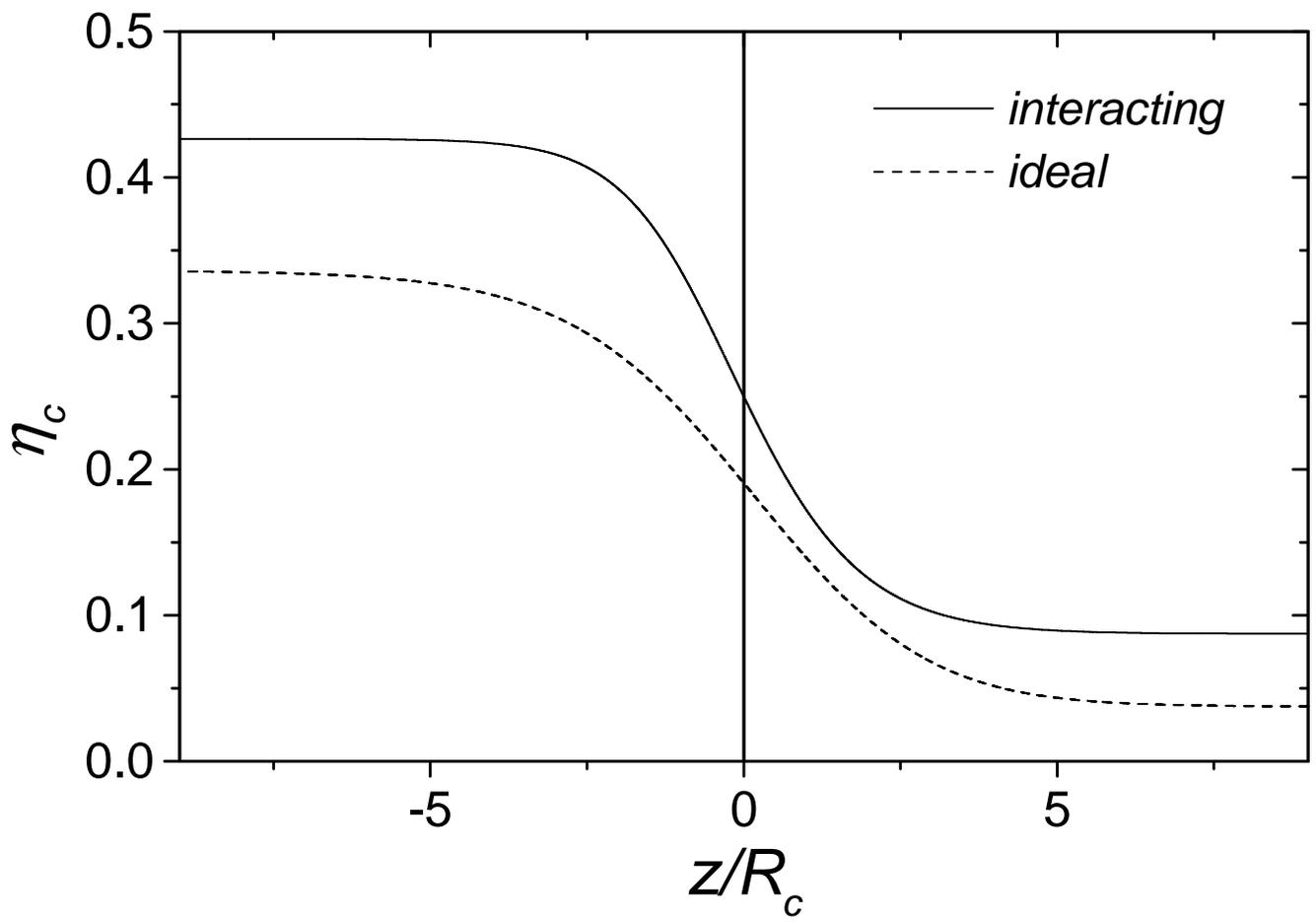}}
\caption{\label{Fig2} Density profiles of the colloidal packing
fraction for ideal and interacting polymers. In both cases, the size
ratio is \(q=1.05\) and the relative polymer reservoir packing
fraction from the critical point is
\((\eta_p^r-\eta_p^{r,crit})/\eta_p^{r,crit}=0.2\). The polymer excluded 
volume interactions yield sharper interfaces.
}
\end{figure}

\newpage

A. Moncho-Jord\'{a} et al., Fig. 3.

\begin{figure}\center\resizebox{1.0\textwidth}{!}{\includegraphics{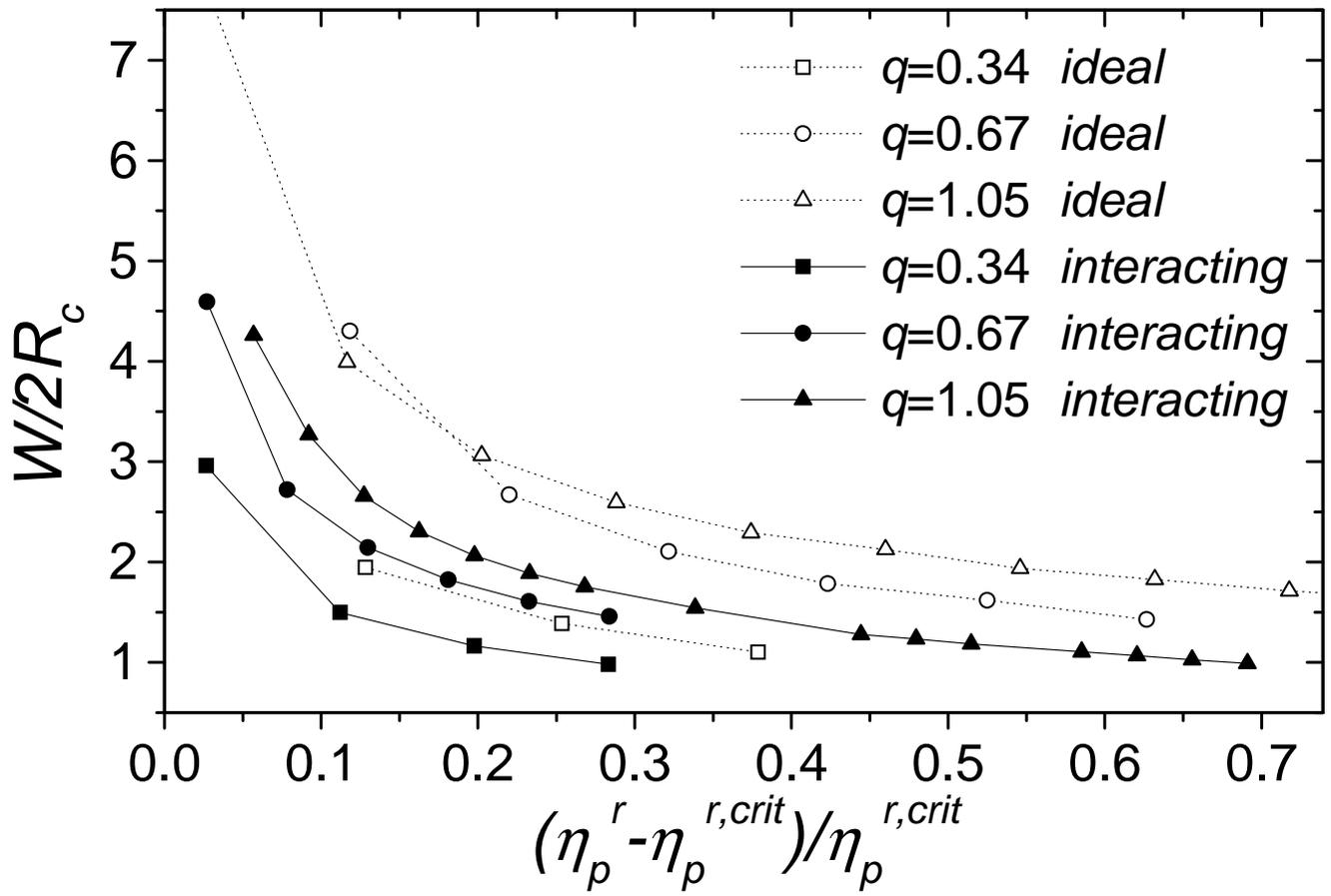}}
\caption{\label{Fig3} 10-90 width of the interface for ideal (white
symbols) and interacting polymers (black symbols) versus the deviation
of the polymer reservoir packing fraction from the critical point, for
\(q=0.34\), 0.67 and 1.05.}
\end{figure}

\newpage

A. Moncho-Jord\'{a} et al., Fig. 4.

\begin{figure} 
\center\resizebox{1.0\textwidth}{!}{\includegraphics{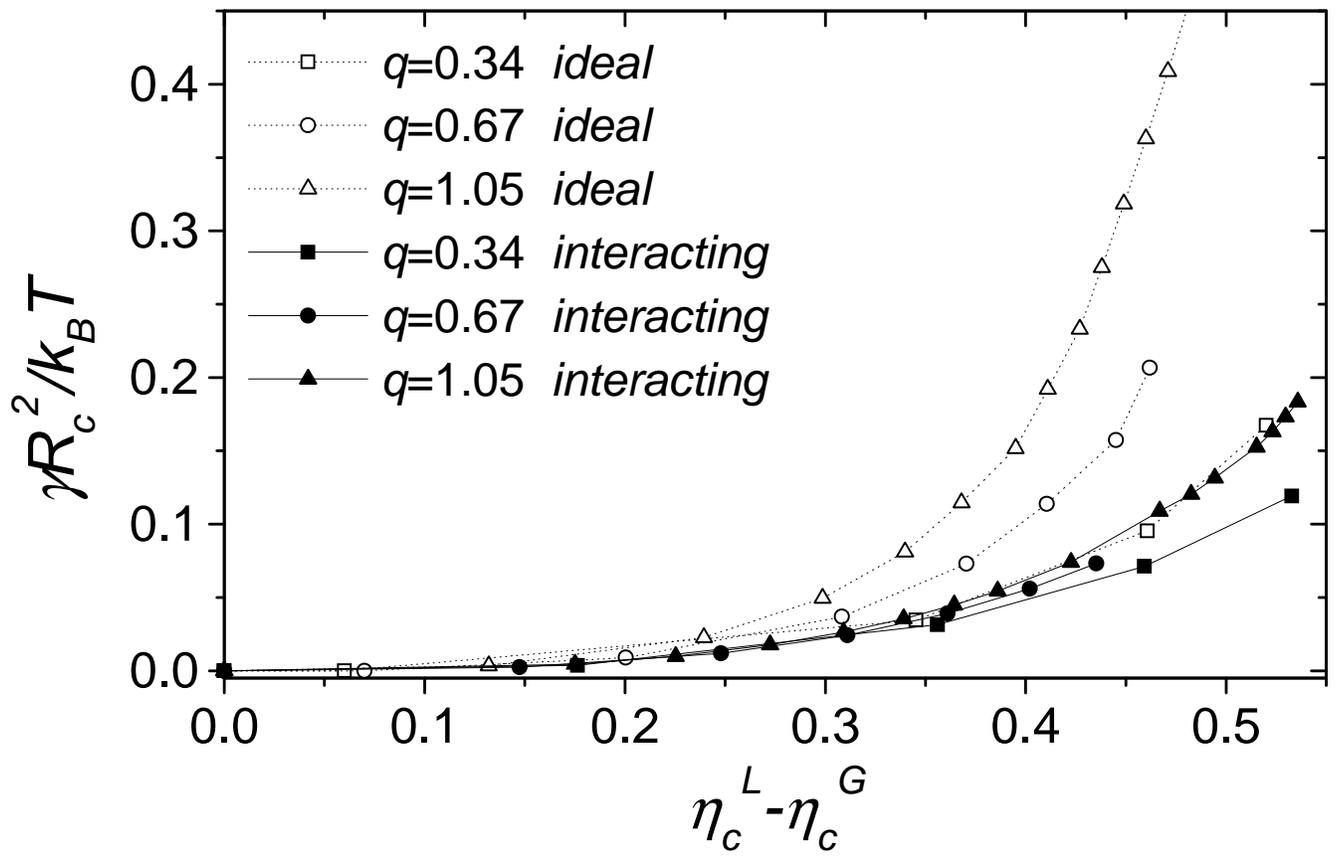}} 
\caption{\label{Fig4} Dimensionless surface tensions for ideal (white 
symbols) and interacting polymers (black symbols) as a function of 
\(\eta_c^L-\eta_c^G\) for \(q=0.34\), 0.67 and 1.05.}
\end{figure} 

\newpage

A. Moncho-Jord\'{a} et al., Fig. 5.

\begin{figure} 
\center\resizebox{1.0\textwidth}{!}{\includegraphics{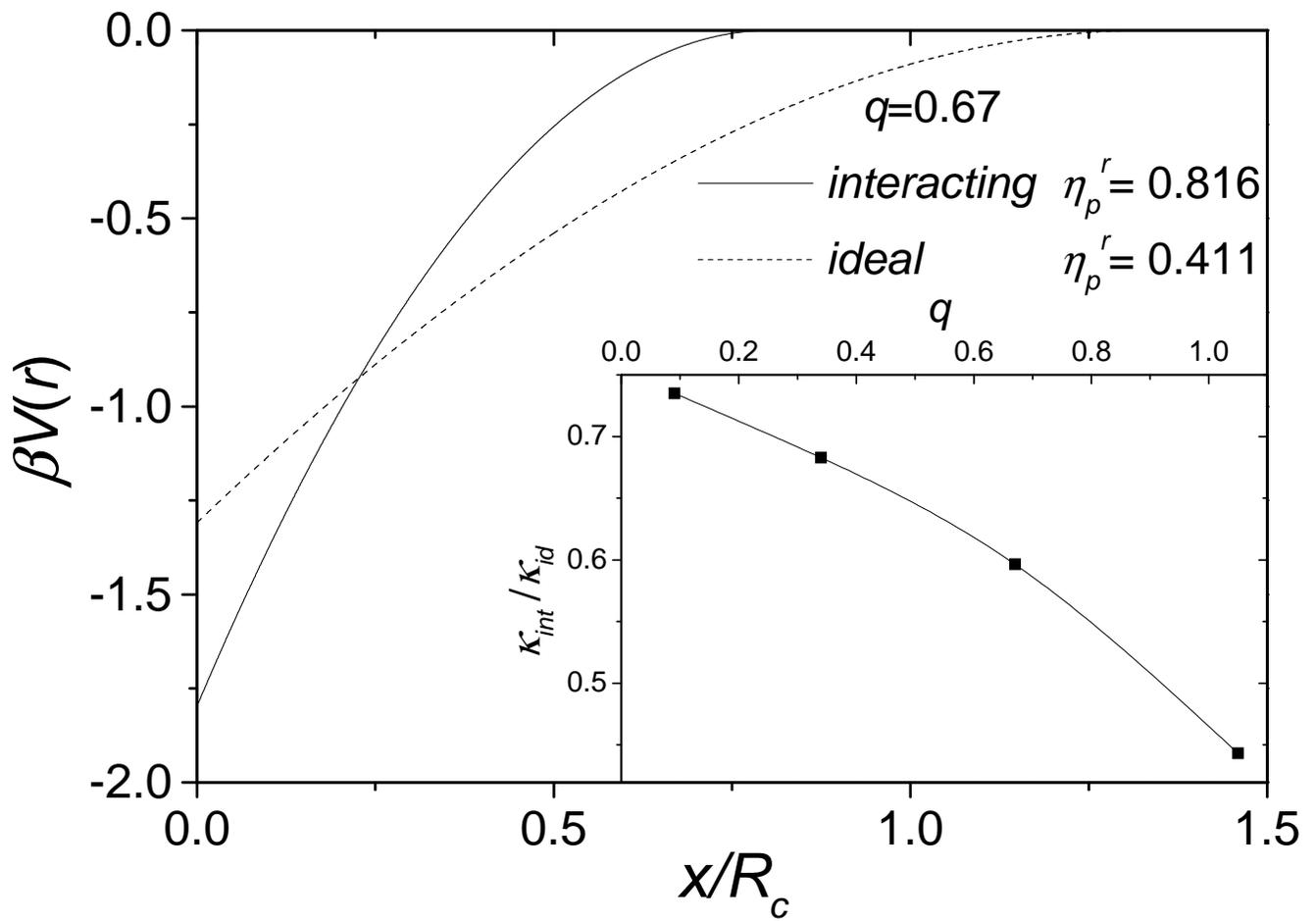}}
\caption{\label{Fig5} Effective colloid-colloid depletion pair 
potential induced by interacting (solid line) and ideal polymers 
(dashed line) for \(q=0.67\) at their respective critical points, as a 
function of the distance between the particle surfaces 
\(x=r-2R_c\). Inset: Ratio between the \(\kappa\) parameter for 
interacting and ideal polymers at the critical point versus the size 
ratio \(q\).}
\end{figure}

\end{document}